\documentclass[british,reprint, aps, onecolumn, final, nofootinbib, superscriptaddress]{revtex4-2}
\PassOptionsToPackage{obeyFinal}{todonotes}
\usepackage[T1]{fontenc}
\usepackage[latin9]{inputenc}
\usepackage{color}
\usepackage{array}
\usepackage{textcomp}
\usepackage{multirow}
\usepackage{todonotes}
\usepackage{amsbsy}
\usepackage{amstext}
\usepackage{amssymb}
\usepackage{graphicx}
\usepackage{esint}

\makeatletter

\providecommand{\tabularnewline}{\\}

\usepackage{xcolor}
\newcommand{\chng}[1]{#1}
\newcommand{\chngb}[1]{#1}

\newcommand{\chngd}[1]{#1}
\newcommand{\deld}{}
\newcommand{\chnge}[1]{#1}
\newcommand{\chngf}[1]{#1}
\newcommand{\chngg}[1]{#1}

\newcommand{\delg}{}

\makeatother

\usepackage{babel}
\begin{document}
\global\long\def\mpl{m_{\mathrm{Pl}}}%
 
\global\long\def\eh{\epsilon_{H}}%
 
\global\long\def\ehc{\epsilon_{H\mathrm{c}}}%
 
\global\long\def\wm{w_{\mathrm{m}}}%
\global\long\def\xc{x_{\mathrm{c}}}%
\global\long\def\yc{y_{\mathrm{c}}}%
\global\long\def\zc{z_{\mathrm{c}}}%
\global\long\def\uc{u_{\mathrm{c}}}%
 
\global\long\def\wf{w_{\mathrm{f}}}%
 
\global\long\def\cgw{c_{\mathrm{GW}}}%
\global\long\def\at{\alpha_{\mathrm{T}}}%

\title{Gauss-Bonnet Dark Energy and the Speed of Gravitational Waves}
\author{José Jaime Terente Díaz}
\affiliation{Departamento de F\'{\i}sica Teórica, Universidad Complutense de Madrid,
E-28040 Madrid, Spain}
\author{Konstantinos Dimopoulos}
\affiliation{Consortium for Fundamental Physics, Physics Department, Lancaster
University, Lancaster LA1 4YB, UK}
\author{Mindaugas Kar\v{c}iauskas}
\affiliation{Departamento de F\'{\i}sica Teórica, Universidad Complutense de Madrid,
E-28040 Madrid, Spain}
\author{Antonio Racioppi}
\affiliation{National Institute of Chemical Physics and Biophysics, Rävala 10,
10143 Tallinn, Estonia}
\begin{abstract}
\chngd{Gauss-Bonnet Dark Energy has been a popular model to explain
the accelerated expansion of the Universe. Quite generically it also
predicts the speed of gravitational waves $\cgw$ to be different
from the speed of light. This fact alone led some authors to exclude
such models in view of the new tight observational constraints on
$\cgw$. However, the behaviour of $\cgw$ depends on the choice of
the Gauss-Bonnet (GB) coupling function. It is possible to construct
models where $\cgw$ is always equal to the speed of light. More generally,
$\cgw$ is a time dependent function with instances where both speeds
coincide. Nevertheless, we observe that the bound on $\cgw$ excludes
scenarios where the GB term directly affects the expansion of the
Universe, even if the constraint on the variation of the coupling
function does not appear to be strong. We perform the dynamical systems
analysis to see if the expansion of the Universe could be affected
indirectly by modulating the behaviour of the scalar field, which
modulates the GB coupling. It is shown that either the bounds on $\cgw$
are violated by many orders of magnitude, or it might be very difficult
to find models that are consistent with other cosmological observations.}
\end{abstract}
\maketitle

\section{Introduction}

The detection of gravitational waves (GW) \citep{LIGOScientific:2016aoc,Abbott:2016nmj,LIGOScientific:2017vwq}
opens a new window to \chngd{observe and} measure the Universe.
Most directly, it enables testing General Relativity (GR) in regimes
that were not accessible before and constrain possible modifications
of the laws of gravity. They also provide new ways to test Dark Energy
(DE) models. Many of such models rely on gravity modifications and
therefore are subject to such constraints.

A very clear demonstration is provided in Ref.~\citep{LIGOScientific:2017zic}.
A lucky coincidence of being able to detect GW emitted by the merger
of two neutron stars as well as the electromagnetic counterpart of
this event made it possible to put very stringent constraints on the
speed of GW, $\cgw$. The delay between arrival times of GW and $\gamma$-rays
led to the bound
\begin{eqnarray}
\left|\at\right| & < & 10^{-15}\,,\label{=0003B1T-bound}
\end{eqnarray}
where $\at$ parametrises the deviation of $\cgw$ from the speed
of light
\begin{eqnarray}
\at & \equiv & \cgw^{2}-1\:,\label{=0003B1T-def}
\end{eqnarray}
\chngd{in} natural units, where $c=\hbar=1$.

Many classes of modified gravity theories predict $\at\ne0$. The
constraints on $\at$ in Ref.~\citep{LIGOScientific:2017zic} excluded
a lot of well motivated and otherwise attractive models and considerably
narrowed down the space of available modifications \citep{Ezquiaga:2017ekz,Amendola:2017ovw}. 

Among the excluded models -- it is claimed in Ref.~\citep{Ezquiaga:2017ekz}
-- is the Gauss-Bonnet Dark Energy (GBDE) one. This model has many
attractive features. The Gauss-Bonnet term itself is a unique combination
of curvature terms squared
\begin{equation}
\mathcal{G}\equiv R^{2}-4R_{\mu\nu}R^{\mu\nu}+R_{\mu\nu\rho\sigma}R^{\mu\nu\rho\sigma}\,,\label{gb-def}
\end{equation}
where $R$, $R_{\mu\nu}$ and $R_{\hphantom{\rho}\sigma\mu\nu}^{\rho}$
are the Ricci scalar, tensor and Riemann tensor respectively. Nevertheless,
this combination leads to metric tensor equations of motion that are
second order. The Gauss-Bonnet (GB) term is quite ubiquitous in actions
of low-energy effective string theory, be it at tree or one loop level
\citep{Callan:1969sn,Gross:1986mw,Boulware:1985wk,Antoniadis:1992rq,Antoniadis:1993jc,Bento:1995qc,Sami:2005zc}.
The corresponding modification can be written as $\xi\mathcal{G}$
term in the Lagrangian, where $\xi$ is the GB coupling. If the latter
is a constant, the GB term is a surface term and can be integrated
out (although it can still be important for other aspects of the theory,
such as regularization \citep{Aros:1999id,Olea:2005gb}). However,
on quite generic grounds, one might expect that the GB term also couples
to scalar fields of the theory, such as moduli or dilaton fields,
making $\xi$ field dependent.

The possibility of explaining DE with the GB term was first investigated
in \chnge{Refs.~\citep{Nojiri:2005vv,Amendola:2005cr,Koivisto:2006ai,Koivisto:2006xf,vandeBruck:2017voa}}.
\chngd{One of the attractive features of such models is that they
provide the means to safely cross the phantom divide, that is, enter
the regime where the DE equation of state is $w<-1$, without instabilities.
The best fit value of $w$ is smaller than $-1$ \citep{Planck:2018vyg}.}
If such $w$ is associated with a scalar field, it leads to many instabilities
and such a model is likely excluded by observations \citep{Cline:2003gs}.
On the other hand, modifications of gravity in GBDE in some parameter
range allow for $w<-1$. In this model $w$ is time dependent\chngd{;
it} briefly dips below $-1$ before settling on $w=-1$ \citep{Koivisto:2006ai}.
Hence, it can accommodate this low value without leading to contradictions.

\chngd{Quite} generically GBDE predicts $\at\ne0$. This fact alone
led the authors of Ref.~\citep{Ezquiaga:2017ekz} to claim that GBDE
is ruled out. \chngd{Here} we would like to point out that $\cgw$,
predicted by GBDE, is not a constant. \chngd{Moreover,} the constraint
in eq.~(\ref{=0003B1T-bound}) is an upper bound which is applicable
only at the very latest stages of the evolution of the Universe.\footnote{\chngg{It is also worth noting that the bound on $\at$ applies only
to the limited range of GW frequencies. This fact alone could save
many Horndeski type Dark Energy models \citep{deRham:2018red}.}} Hence, to assess the implications of these constraints for GBDE we
need to study it more carefully.

In this work we use the dynamical \chngd{systems} analysis to look
for viable models of GBDE and compute the evolution of $\at$ parameter.
The crucial quantity in such models is the GB coupling $\xi\left(\phi\right)$.
It determines the dynamics of the universe as well as the evolution
of the $\at$ parameter. \chngd{Applying} the bound in eq.~(\ref{=0003B1T-bound})
to the variation of $\xi\left(\phi\right)$ and the rate of its variation,
we find that the \chngd{constraints appear} weak. \chngd{Nevertheless,
the bound in eq.~(\ref{=0003B1T-bound}) prevents the GB term from
affecting the expansion of the Universe directly. The remaining possibility
is for this term to affect the expansion indirectly, by modifying
the behaviour of the scalar field. To investigate this issue we apply
the dynamical systems analysis. We also apply this analysis to the
case where $\at=0$ by construction, which is allowed by the model.}

In Section~\ref{sec:dyn-sys} we introduce the model, derive dynamical
equations and show the bounds on $\xi\left(\phi\right)$ that follow
from eq.~(\ref{=0003B1T-bound}). In Section~\ref{sec:dynless}
we assume an exponential potential $V\left(\phi\right)$ and \deld
write the dynamical equations in terms of dimensionless variables,
which are used \chngd{in the following sections. The dynamical systems
analysis is applied to models with the exponential GB coupling $\xi\left(\phi\right)$
in Section~\ref{sec:Exponential} and it is applied to the linear
function $\xi\left(\phi\right)$ in Section~\ref{sec:lin}. The case
of $\at=0$ is studied in Section~\ref{sec:aT=00003D0}.}

\section{Scalar-Gauss-Bonnet Dark Energy and Constraints on $\protect\cgw$\label{sec:dyn-sys}}

We start with the scalar-Gauss-Bonnet action 
\begin{eqnarray}
S & = & \int\mathrm{d}^{4}x\sqrt{-g}\left[\frac{1}{2}\mpl^{2}R+\xi\left(\phi\right)\mathcal{G}-\frac{1}{2}\partial_{\mu}\phi\partial^{\mu}\phi-V\left(\phi\right)+\mathcal{L}_{\mathrm{m}}\right]\,,\label{S}
\end{eqnarray}
where $\mathcal{G}$ is defined in eq.~(\ref{gb-def}). For brevity
we address to the above action as the Gauss-Bonnet (GB) action in
this work. $\mathcal{L}_{\mathrm{m}}$ is the Lagrangian of the matter
sector. If $\xi$ is constant, the GB term is a total derivative and
does not affect the dynamics of the system. We assume the background
spacetime to be homogeneous, isotropic and flat, described by the
FRW metric $g_{\mu\nu}=\mathrm{diag}\left[-1,a^{2}\left(t\right),a^{2}\left(t\right),a^{2}\left(t\right)\right]$,
where $t$ is the cosmic time and $a$ is the scale factor.

In this model the speed of tensor mode propagation is determined by
the rate of change of the coupling function $\xi\left(\phi\right)$
\citep{Noh:2001ia,Hwang:2005hb,Hwang:2002fp,Odintsov:2019clh} (see
also Refs.~\citep{Kawai:1998ab,Kawai:1999pw})
\begin{eqnarray}
\alpha_{\mathrm{T}} & = & \frac{8\left(\ddot{\xi}-\dot{\xi}H\right)}{\mpl^{2}+8\dot{\xi}H}\,,\label{=0003B1T}
\end{eqnarray}
where we use a dot to denote the derivative with respect to $t$ and
$H\equiv\dot{a}/a$ is the Hubble parameter. It is clear from this
expression that the constraint in eq.~(\ref{=0003B1T-bound}) can
be satisfied if one of the following two conditions is fulfilled.
The first option is to choose the coupling function $\xi\left(\phi\right)$
such that
\begin{eqnarray}
\ddot{\xi} & = & H\dot{\xi}\,.\label{bound-spec}
\end{eqnarray}
This choice is discussed in refs.~\citep{Odintsov:2019clh,Oikonomou:2022xoq}
(other related references can be found in these articles) in the context
of inflation. However, there is another, more generic possibility.
We notice in eq.~(\ref{=0003B1T}) that $\alpha_{T}$ is suppressed
by the Planck mass. Therefore, as long as the conditions $\left|\ddot{\xi}\right|/\mpl^{2},\:H\left|\dot{\xi}\right|/\mpl^{2}<10^{-15}$
are satisfied, the GB action in eq.~(\ref{S}) is compatible with
the constraints on the speed of gravitational waves. We can write
these conditions in a more useful way
\begin{eqnarray}
\frac{\left|\ddot{\xi}\right|}{H^{2}},\,\frac{\left|\dot{\xi}\right|}{H} & < & 10^{-15}\left(\frac{\mpl}{H}\right)^{2}\,,\label{bound-gen}
\end{eqnarray}
which emphasises the change of the coupling function and the rate
of change of this coupling over one Hubble time. If this condition
is to be imposed on\chngf{ large field inflation models, where $H\lesssim10^{-5}\mpl$
\citep{Planck:2018jri}}, this bound can be tight. In that case,
to limit $\alpha_{T}$ within the allowed range, it is better to look
for models that satisfy the condition in eq.~(\ref{bound-spec}).
However, the constraints from the observations of GRB170817A \chngf{do
not apply to the early Universe. Therefore it is not very useful to
use them in that context. Instead, this constraint is applicable to
the present Universe, }within a very narrow range of e-folds (see
Fig.~\ref{fig:=0003B1T}), where the Hubble parameter is more than
fifty orders of magnitude smaller than \chngf{the value of $H$ cited
above}. Indeed, plugging $H_{0}^{2}/\mpl^{2}\sim10^{-120}$ \citep{Planck:2018vyg}
into eq.~(\ref{bound-gen}) we find\footnote{\chngf{This is only an order of magnitude estimate. Factors of order
1 do not change the conclusions in any substantial way.}}
\begin{eqnarray}
\frac{\left|\ddot{\xi}\right|}{H_{0}^{2}},\,\frac{\left|\dot{\xi}\right|}{H_{0}} & < & 10^{105}\,.\label{bound-gen-num}
\end{eqnarray}
That is, $\xi$ \chngb{and $\dot{\xi}$ need to vary by more than
100 orders of magnitude, $\Delta\xi\:,\Delta\dot{\xi}<10^{105}$,}
over the age of the Universe to violate the bound. \deld This appears
to demonstrate that the constraint on $c_{\mathrm{GW}}$ is exceptionally
weak and might give hope that GBDE models remain viable. Unfortunately,
as we show in this work, \chngd{at least for simple functions $\xi\left(\phi\right)$,}
this turns out not to be the case.

To understand the implications of eq.~(\ref{=0003B1T-bound}) for
Gauss-Bonnet Dark Energy (GBDE) models better, let us first write
the homogeneous dynamical equations in the FRW background as \citep{Nojiri:2005vv,Heydari-Fard:2016nlj}
\begin{eqnarray}
H^{2} & = & \frac{\rho_{\phi}+\rho_{\mathrm{m}}}{3\left(\mpl^{2}+8H\dot{\xi}\right)}\:,\label{H2}\\
\dot{H} & = & -\frac{\rho_{\phi}+P_{\phi}+\rho_{\mathrm{m}}+P_{\mathrm{m}}+8H^{2}\left(\ddot{\xi}-\dot{\xi}H\right)}{2\left(\mpl^{2}+8H\dot{\xi}\right)}\:,\label{dH}
\end{eqnarray}
where $\rho_{\phi}$ and $P_{\phi}$ are the energy and pressure densities
of the homogeneous scalar field $\phi$ respectively. They are defined
to be
\begin{eqnarray}
\rho_{\phi} & \equiv & \frac{1}{2}\dot{\phi}^{2}+V\left(\phi\right)\:,\label{=0003C1=0003D5}\\
P_{\phi} & \equiv & \frac{1}{2}\dot{\phi}^{2}-V\left(\phi\right)\:.\label{P=0003D5}
\end{eqnarray}
Similarly $\rho_{\mathrm{m}}$ and $P_{\mathrm{m}}$ are the energy
and pressure densities of the matter field.

The acceleration of spatial slices can be parametrised using the Hubble
flow parameter
\begin{eqnarray}
\epsilon_{H} & \equiv & -\frac{\dot{H}}{H^{2}}\:.\label{eH}
\end{eqnarray}
Alternatively, it is common to use the deceleration parameter $q\equiv\eh-1$
for this purpose. The spatial slices expand in an accelerating fashion
if $\eh<1$ ($q<0$).

Plugging eqs.~(\ref{H2}) and (\ref{dH}) into eq.~(\ref{eH}) we
can write
\begin{eqnarray}
\eh & = & \frac{3}{2}\left(1+\frac{P_{\phi}+P_{\mathrm{m}}}{\rho_{\phi}+\rho_{\mathrm{m}}}\right)+\frac{1}{2}\alpha_{T}\:,\label{=0003F5H-GB}
\end{eqnarray}
where we also made use of eq.~(\ref{=0003B1T}). At the present epoch
$\eh\simeq0.5$. Thus, in view of eq.~(\ref{=0003B1T-bound}) we
see that the last term must be negligible. This rules out any direct
effect of the Gauss-Bonnet term to the expansion of the Universe.
Neglecting that last term, we arrive at the expression which can also
be obtained in a typical, General Relativistic quintessence models
\citep{Ratra:1987rm,Wetterich:1987fm}.

But even if observations exclude the scenario where the GB term affects
the expansion of the Universe directly, there remains a possibility
that it does so indirectly, by modifying the behaviour of the scalar
field $\phi$. As we will see next, such a possibility is also excluded,
at least for an exponential potential $V\left(\phi\right)$.

\section{The Dynamical System\label{sec:dynless}}

Equations (\ref{H2}) and (\ref{dH}) can be supplemented with dynamical
equations governing the evolution of the $\phi$ field and $\rho_{\mathrm{m}}$
\begin{eqnarray}
\ddot{\phi}+3H\dot{\phi}+V_{,\phi} & = & 24\xi_{,\phi}\left(\dot{H}+H^{2}\right)H^{2}\:,\label{phi-eom}\\
\dot{\rho}_{\mathrm{m}}+3H\rho_{\mathrm{m}}\left(1+\wm\right) & = & 0\:,\label{rhom-cont}
\end{eqnarray}
where $\wm\equiv P_{\mathrm{m}}/\rho_{\mathrm{m}}$ is the barotropic
parameter of the matter component. And we assume a matter fluid with
$0\le\wm<1$. 

To analyse the generic behaviour of this dynamical system, it is convenient
to normalise the dynamical degrees of freedom and write them in a
dimensionless form, such as
\begin{equation}
x\equiv\frac{\phi'}{\sqrt{6}\mpl}\,,\quad y\equiv\frac{\sqrt{V}}{\sqrt{3}\mpl H}\:,\quad u\equiv4\sqrt{6}\frac{H^{2}\xi_{,\phi}}{\mpl}\,,\text{ and }z\equiv\frac{\sqrt{\rho_{\mathrm{m}}}}{\sqrt{3}\mpl H}\,.\label{dless-def}
\end{equation}
The prime in the definition of $x$ and the equations below denotes
the derivatives with respect to the e-fold number
\begin{eqnarray}
N & \equiv & \ln a\,,
\end{eqnarray}
where we normalised $a$ such that $a=1$ today. \chngf{The above
defined dimensionless parameters are related to the more familiar
density parameters by
\begin{eqnarray}
\Omega_{\mathrm{m}} & \equiv & \frac{\rho_{\mathrm{m}}}{3\mpl^{2}H^{2}}=z^{2}\:,\label{=0003A9m}\\
\Omega_{\phi} & \equiv & \frac{\rho_{\phi}}{3\mpl^{2}H^{2}}=x^{2}+y^{2}\:,\\
\Omega_{\mathrm{GB}} & \equiv & -\frac{8H\dot{\xi}}{\mpl^{2}}=-2ux\:.\label{=0003A9GB}
\end{eqnarray}
The first two definitions are exactly the same as in models of GR
with a scalar field. The physical origin of the last parameter is
due to the modifications of gravity, but it is interpreted as an effective
matter fluid. Following this interpretation we can write the constraint
equation (\ref{H2}) (the Friedmann equation) as
\begin{equation}
1=\Omega_{\phi}+\Omega_{\mathrm{m}}+\Omega_{\mathrm{GB}}\:.\label{=0003A9const}
\end{equation}
In GR the analogous equation confines the range of variation of each
parameter to $\left|\Omega\right|\le1$. This is due to the density
parameters being non-negative. In GB gravity $\Omega_{\mathrm{GB}}$
can be positive as well as negative. This makes it possible for $\Omega_{\phi}$
and $\Omega_{\mathrm{m}}$ to exceed unity.}

The definitions in eq.~(\ref{dless-def}) are particularly useful
if the scalar field potential is an exponential function
\begin{eqnarray}
V & = & V_{0}\mathrm{e}^{-\lambda\phi/\mpl}\:,\label{V-spec}
\end{eqnarray}
where we take $\lambda>0$ to be a constant. We will always use the
above ansatz in this work. In that case the dynamical equations are
self-similar and the explicit dependence on the Hubble parameter drops
out of those equations. In particular, eqs.~(\ref{phi-eom}) and
(\ref{rhom-cont}) can be written as
\begin{eqnarray}
x' & = & \left(\eh-3\right)x+\sqrt{\frac{3}{2}}\lambda y^{2}+u\left(1-\eh\right)\:,\label{dx}\\
y' & = & \left(\eh-\sqrt{\frac{3}{2}}\lambda x\right)y\:,\label{dy}\\
z' & = & \left[\eh-\frac{3}{2}\left(1+\wm\right)\right]z\:,\label{dz}
\end{eqnarray}
where
\begin{eqnarray}
\eh & = & \left[3x^{2}+\frac{3}{2}\left(1+\wm\right)z^{2}+\left(ux\right)'-ux\right]\frac{1}{1+ux}\label{eH-dyn}
\end{eqnarray}
\chngf{The constraint equation (\ref{=0003A9const}), in terms of
the dimensionless variables, can be written as
\begin{eqnarray}
1 & = & x^{2}+y^{2}+z^{2}-2ux\:.\label{constr}
\end{eqnarray}

When doing dynamical analysis of this system, it is convenient to
use the equation for $u$ too. Taking the derivative of the expression
in eq.~(\ref{dless-def}) we find
\begin{eqnarray}
u' & = & -2\eh u+24H^{2}\xi_{,\phi\phi}\:x\:.\label{du}
\end{eqnarray}

A few comments about these equations are in order. First, notice that
taking $u=0$ ($\xi=\mathrm{const}$) we recover the same equations
as used for models in General Relativity (e.g. ref.~ \citep{Copeland:1997et}).
In those models, the dimensionless variables $y$ and $z$ are constrained
within the range $\left[0;1\right]$ and $x\in\left[-1;1\right]$.
This can be seen from eq.~(\ref{constr}) or equivalently from eq.~(\ref{=0003A9const}).
In the case of the GB models such restrictions do not apply. As mentioned
above, \emph{a priori} the sign of $u$ (or $\Omega_{\mathrm{GB}}$)
is not determined. Hence, the maximum values of $\left|x\right|$,
$y$ and $z$ are not limited to $1$. Second, eq.~(\ref{eH-dyn})
diverges as $ux\rightarrow-1$. However, this value is never reached
because it falls within the phase space region that is forbidden by
the constraint equation (\ref{constr}).

}

\section{The Exponential Gauss-Bonnet Coupling\label{sec:Exponential}}

\subsection{Dynamics}

To understand the qualitative behaviour of this dynamical system,
we find its fixed points and investigate their stability. The fixed
points are defined as points, or regions, in the phase space where
$x'=y'=z'=0$ is satisfied. First, we study the case of the exponential
GB function, given by
\begin{equation}
\xi=\xi_{0}\mathrm{e}^{\kappa\phi/\mpl}\,,\label{=0003BE=0003BA}
\end{equation}
which allows us to write eq.~(\ref{du}) as
\begin{eqnarray}
u' & = & \left(\sqrt{6}\kappa x-2\eh\right)u\:.\label{du-exp}
\end{eqnarray}
The computation details for finding fixed points are provided in the
\textbf{Appendix} and the results are summarised in Table~\ref{tab:pc-gen}.

\begin{table}

\begin{centering}
\begin{tabular}{|c|c|c|c|c|c|c|c|}
\cline{2-8} \cline{3-8} \cline{4-8} \cline{5-8} \cline{6-8} \cline{7-8} \cline{8-8} 
\multicolumn{1}{c|}{} & $\xc$ & $\yc$ & $\uc$ & $\zc$ & Expansion rate $\ehc$ & If $\kappa$ is & $\ehc\uc$\tabularnewline
\hline 
\textbf{M} & 0 & 0 & 0 & 1 & $\frac{3}{2}\left(1+\wm\right)$ & \multirow{5}{*}{any $\kappa$} & \multirow{5}{*}{0}\tabularnewline
\cline{1-6} \cline{2-6} \cline{3-6} \cline{4-6} \cline{5-6} \cline{6-6} 
\textbf{K\textpm{}} & $\pm1$ & 0 & 0 & 0 & $3$ &  & \tabularnewline
\cline{1-6} \cline{2-6} \cline{3-6} \cline{4-6} \cline{5-6} \cline{6-6} 
\textbf{I} & $\frac{\lambda}{\sqrt{6}}$ & $\sqrt{1-\frac{\lambda^{2}}{6}}$ & 0 & 0 & $\frac{1}{2}\lambda^{2}$ &  & \tabularnewline
\cline{1-6} \cline{2-6} \cline{3-6} \cline{4-6} \cline{5-6} \cline{6-6} 
\textbf{Sc} & $\sqrt{\frac{3}{2}}\frac{1+\wm}{\lambda}$ & $\frac{\sqrt{\frac{3}{2}\left(1-\wm^{2}\right)}}{\lambda}$ & 0 & $\sqrt{1-\frac{3\left(1+\wm\right)}{\lambda^{2}}}$ & $\frac{3}{2}\left(1+\wm\right)$ &  & \tabularnewline
\cline{1-6} \cline{2-6} \cline{3-6} \cline{4-6} \cline{5-6} \cline{6-6} 
\textbf{dS} & 0 & $1$ & $-\sqrt{\frac{3}{2}}\lambda$ & 0 & 0 &  & \tabularnewline
\hline 
\textbf{S2} & $\sqrt{\frac{3}{2}}\frac{1+\wm}{\kappa}$ & 0 & $\sqrt{\frac{3}{2}}\frac{3\left(\wm^{2}-1\right)}{\kappa\left(3\wm+1\right)}$ & $\sqrt{1+\frac{3\left(3\wm-7\right)\left(1+\wm\right)^{2}}{2\kappa^{2}\left(3\wm+1\right)}}$ & $\frac{3}{2}\left(1+\wm\right)$ & \multirow{2}{*}{$\kappa\ne\lambda$} & \multirow{4}{*}{$\ne0$}\tabularnewline
\cline{1-6} \cline{2-6} \cline{3-6} \cline{4-6} \cline{5-6} \cline{6-6} 
\textbf{G} & $\beta$ & 0 & $\frac{\beta^{2}-1}{2\beta}$ & 0 & $\frac{5\beta^{2}+1}{\beta^{2}+1}=\sqrt{\frac{3}{2}}\kappa\beta$ &  & \tabularnewline
\cline{1-7} \cline{2-7} \cline{3-7} \cline{4-7} \cline{5-7} \cline{6-7} \cline{7-7} 
\textbf{S3} & $\sqrt{\frac{3}{2}}\frac{1+\wm}{\lambda}$ & $\frac{\sqrt{\frac{3}{2}\left(1-\wm^{2}\right)+\frac{\lambda}{\sqrt{6}}\left(1+3\wm\right)\uc}}{\lambda}$ & $\uc$ & $\sqrt{1-\frac{3\left(1+\wm\right)-\frac{\lambda}{\sqrt{6}}\left(5+3\wm\right)\uc}{\lambda^{2}}}$ & $\frac{3}{2}\left(1+\wm\right)$ & \multirow{2}{*}{$\kappa=\lambda$} & \tabularnewline
\cline{1-6} \cline{2-6} \cline{3-6} \cline{4-6} \cline{5-6} \cline{6-6} 
\textbf{IV} & $\xc$ & $\sqrt{1-\xc^{2}+2\xc\cdot\frac{3\xc-\sqrt{\frac{3}{2}}\lambda}{1+\sqrt{\frac{3}{2}}\lambda\xc}}$ & $\frac{3\xc-\sqrt{\frac{3}{2}}\lambda}{1+\sqrt{\frac{3}{2}}\lambda\xc}$ & 0 & $\sqrt{\frac{3}{2}}\lambda\xc$ &  & \tabularnewline
\hline 
\end{tabular}
\par\end{centering}

\caption{\label{tab:pc-gen}The values of $x$, $y$, $u$ and $z$ parameters
at the fixed points and curves. The conditions for the existence of
these fixed points and curves can be derived from the fact that all
$\protect\xc$, $\protect\yc$, $\protect\uc$ and $\protect\zc$
values must be real. Parameters $\lambda$, $\kappa$ and $\beta$
are defined in eqs.~(\ref{V-spec}), (\ref{=0003BE=0003BA}) and
(\ref{beta-def}) respectively. Both $\lambda$ and $\kappa$ are
strictly positive.}
\end{table}

Looking at this table, notice that all the fixed points with $\uc=0$
coincide with the ones analysed in ref.~\citep{Copeland:1997et},
as expected. In this reference the authors analyse a quintessence
model within the theory of GR and the exponential scalar field potential.
In other words, all the fixed points that are present in a similar
setup in GR, they are also present in GB models. However, even if
the fixed points coincide, the presence of the GB term might change
their stability, as we will show below.

The fixed point \textbf{M} in Table~\ref{tab:pc-gen} corresponds
to the case where the scalar field is diluted and only the matter
field remains. The \textbf{K\textpm{}} fixed points correspond to
kination, where the universe is dominated by the kinetic energy of
the scalar field. In the case of \textbf{I} and $\lambda<\sqrt{2}$
this fixed point represents the power law inflation \citep{Lucchin:1984yf}.
In the scaling fixed point (\textbf{Sc}) the evolution of the scalar
field adjusts to mimic the behaviour of the matter field. Therefore,
the expansion rate of the Universe is given by $\eh=\frac{3}{2}\left(1+\wm\right)$.

The GB term introduces two more scaling solutions: the fixed point
\textbf{S2}, which exists for $\kappa\ne\lambda$, and the fixed curve
\textbf{S3}, which exists if $\xi\left(\phi\right)V\left(\phi\right)=\mathrm{constant}$.
For our purpose, the most interesting new fixed point is the de Sitter
one (\textbf{dS}), where $\dot{H}=0$. This fixed point is very robust,
and exists for large variety of $\xi\left(\phi\right)$ functions.

Some discussion of DE models with $\kappa=\lambda$ and various solutions
were provided in Ref.~\citep{Nojiri:2005vv}. Some dynamical analysis
with $\kappa\ne\lambda$ was performed in Ref.~\citep{Koivisto:2006ai}
(see also \citep{Tsujikawa:2006ph}). Here we modify and extend the
analysis to make it more generic. In this section, we take $\kappa\ne\lambda$. 

Often, the stability of fixed points can be determined by taking a
linear perturbation of equations (\ref{dx})--(\ref{du}) around
those points. In the case of an exponential GB function in eq.~(\ref{=0003BE=0003BA})
those linear equations can be written as

\begin{eqnarray}
\delta x' & = & \left(\ehc-3\right)\delta x+\sqrt{6}\lambda\yc\delta y+\left(1-\ehc\right)\delta u+\left(\xc-\uc\right)\delta\eh\,,\label{delx-exp}\\
\delta y' & = & -\sqrt{\frac{3}{2}}\lambda\yc\delta x+\left(\ehc-\sqrt{\frac{3}{2}}\lambda\xc\right)\delta y+\yc\delta\eh\,,\\
\delta u' & = & \sqrt{6}\kappa\uc\delta x+\left(\sqrt{6}\kappa\xc-2\ehc\right)\delta u-2\uc\delta\eh\,,\label{delu-exp}\\
\delta z' & = & \left[\ehc-\frac{3}{2}\left(1+\wm\right)\right]\delta z+\zc\delta\eh\,,\label{delz-exp}
\end{eqnarray}
where
\begin{eqnarray}
\delta\eh & = & \frac{\uc\delta x'+\xc\delta u'+\left[6\xc-\left(\ehc+1\right)\uc\right]\delta x-\xc\left(1+\ehc\right)\delta u+2\ehc\zc\delta z}{1+\uc\xc}
\end{eqnarray}
is the linearised eq.~(\ref{eH-dyn}). The constraint equation fixes
the dynamics onto the three-dimensional hypersurface in the four-dimensional
phase space. The linearised version of that equation is given by 
\begin{eqnarray}
0 & = & \left(\xc-\uc\right)\delta x+\yc\delta y+\zc\delta z-\xc\delta u\:.
\end{eqnarray}

We next compute the eigenvalues of the system of equations (\ref{delx-exp})-(\ref{delz-exp})
and determine their stability. The eigenvalues at the fixed point
\textbf{M} are 
\begin{eqnarray}
m_{1} & = & \ehc-3\:,\\
m_{2} & = & \ehc\:,\\
m_{3} & = & -2\ehc\:,
\end{eqnarray}
where $\ehc=\frac{3}{2}\left(1+\wm\right)$ is the value of the Hubble
flow parameter at \textbf{M}. We can see that this fixed point is
always a saddle point for the range of $\wm$ values that we consider.

The eigenvalues at the kination fixed points (\textbf{K\textpm })
are 
\begin{eqnarray}
m_{1} & = & \frac{3}{2}\left(1-\wm\right)\:,\\
m_{2} & = & -6\pm\sqrt{6}\kappa\:,\\
m_{3} & = & 3\mp\sqrt{\frac{3}{2}}\lambda\:,
\end{eqnarray}
where the upper sign corresponds to the point \textbf{K+}. We can
see that the eigenvalue $m_{1}$, which corresponds to the eigenvector
$\boldsymbol{v}=\left(0,0,0,1\right)$\footnote{We order eigenvector components as $\boldsymbol{v}=\left(v_{x},v_{y},v_{u},v_{z}\right)$.},
is always positive. Hence, these two fixed points are never stable.

The eigenvalues at the scaling fixed point \textbf{Sc} are
\begin{eqnarray}
m_{1} & = & 2\left(\frac{\kappa}{\lambda}-1\right)\ehc\:,\\
m_{\pm} & = & -\frac{1}{2}\left(3-\ehc\right)\left[1\pm\sqrt{1-\frac{8\ehc}{3-\ehc}\cdot\left(1-2\frac{\ehc}{\lambda^{2}}\right)}\right]\:.
\end{eqnarray}
As can be seen from Table~\ref{tab:pc-gen}, this fixed point exists
($\zc^{2}\ge0$) only if $\lambda^{2}>2\ehc$. Such a condition makes
the real part of $m_{\pm}$ always negative. Therefore the stability
of this fixed point is determined solely by the sign of $m_{1}$.
That is, the scaling fixed point \textbf{Sc} is a saddle point for
$\kappa>\lambda$.

The eigenvalues at the de Sitter fixed point \textbf{dS} are
\begin{eqnarray}
m_{1} & = & -\frac{3}{2}\left(1+\wm\right)\:,\\
m_{\pm} & = & -\frac{3}{2}\left[1\pm\sqrt{1+\frac{8\lambda^{2}}{3\left(2+3\lambda^{2}\right)}\left(1-\frac{\kappa}{\lambda}\right)}\right]\:.
\end{eqnarray}
Notice that the condition for the stability of this fixed point is
exactly opposite from the one required by the scaling fixed point
\textbf{Sc}: for $\kappa>\lambda$ the scaling fixed point is a saddle
and the de Sitter one is the attractor. \chngf{That is, only one
point is attractive, either \textbf{dS} or \textbf{Sc}, depending
on the magnitude of $\kappa/\lambda$ ratio. }None of the interesting
solutions pass through the fixed points \textbf{G} or \textbf{S2}
so we don't analyse their stability here.

\begin{figure}
\begin{centering}
\includegraphics[scale=0.4]{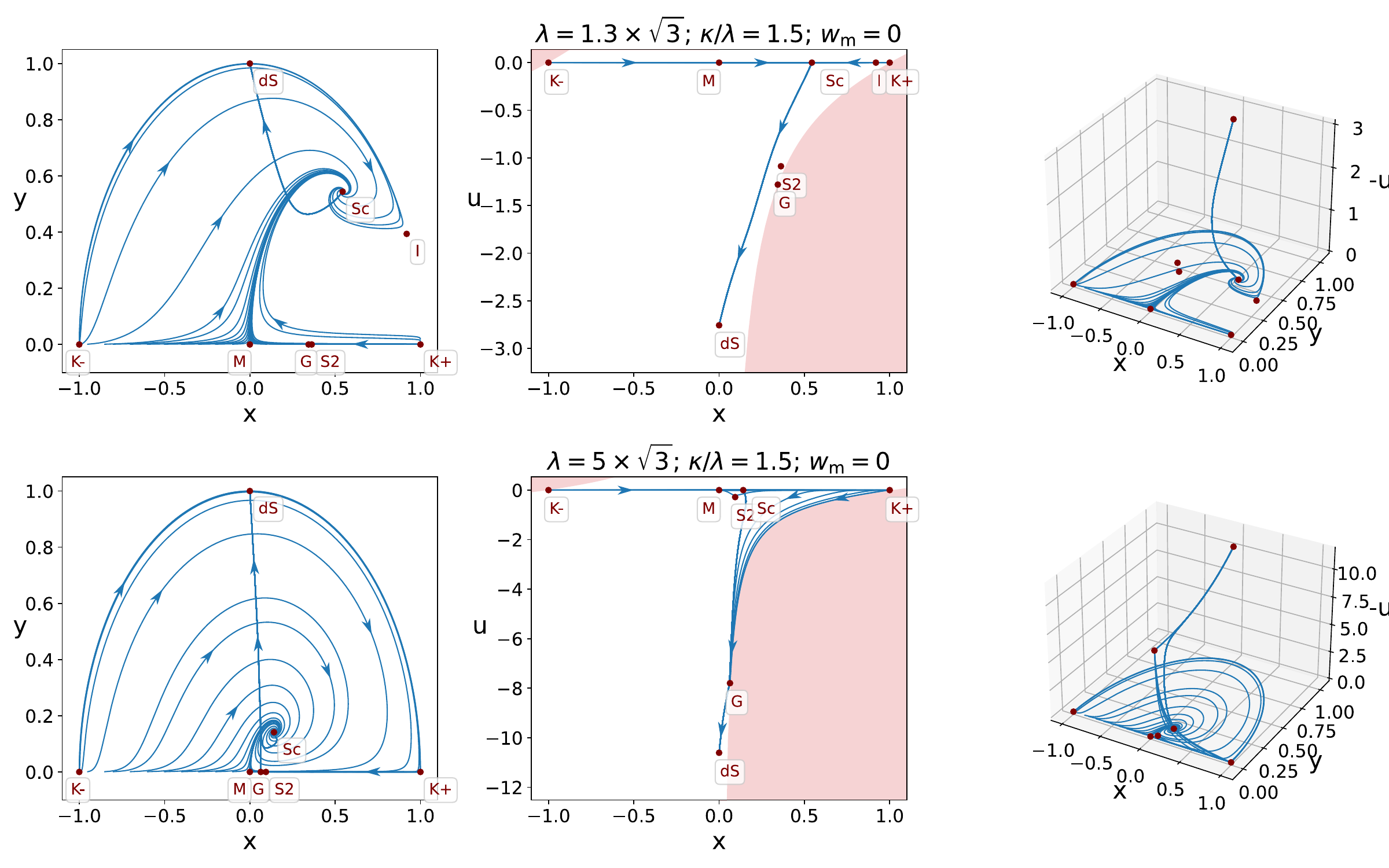}
\par\end{centering}
\begin{centering}
\caption{\label{fig:portraits}The phase portrait of the dynamical system given
by eqs.~(\ref{dx}), (\ref{dy}) and (\ref{du-exp}), with the matter
component represented by a pressureless fluid, $\protect\wm=0$. All
trajectories start with very small values of $y$ and $u$. The first
two columns show a projection of the 3 dimensional phase space. The
full 3D portrait is displayed in the last column. Red regions in the
second column mark forbidden regions of the phase space, where the
constraint equation (\ref{constr}) cannot be satisfied for all values
of $y$ and $z$. For those values the constraint in eq.~(\ref{constr})
cannot be satisfied. The upper plot corresponds to $\lambda=1.3\times\sqrt{3}$
and the lower one to $\lambda=5\sqrt{3}$. In both cases $\kappa/\lambda=3/2$.}
\par\end{centering}
\end{figure}

To visualise the behaviour of the system we integrate numerically
a set of trajectories and show the phase portraits in Figure~\ref{fig:portraits}.
\chngf{ All the trajectories in the plots start from $y_{0}=10^{-3}$
and move towards the \textbf{dS} attractor at $y=1$. We start our
simulations with a negligible GB contribution, $u_{0}=-10^{-25}$
to be precise. A few examples of phase portraits with other values
of $u_{0}$ are shown in Appendix~\ref{sec:Phase-Portraits-u}. Positive
$u_{0}$ values are not viable for our purpose to explain DE. All
such trajectories move towards large values of $u>0$, away from the
de Sitter fixed point \textbf{dS} (see Figure~\ref{fig:phpt-u}).
Another option is to take negative values of $u_{0}$ and $\left|u_{0}\right|\sim\mathcal{O}\left(1\right)$.
As we can see in Figure~\ref{fig:phpt-u}, this option could provide
promising candidate trajectories to explain Dark Energy. This is due
to the fact that a large portion of phase space initially evolves
towards $u\rightarrow0$, reaches the scaling fixed point \textbf{Sc}
and then follows the same evolution pattern as the trajectories with
$\left|u_{0}\right|\ll1$. Nevertheless for our analysis we choose
$\left|u_{0}\right|\ll1$, as this increases the parameter space for
viable candidate trajectories. Such small values are also consistent
with the scenario proposed in ref.~\citep{vandeBruck:2017voa}, where
the GB contribution is negligible initially.}

The physically interesting trajectories are those that start close
to the \textbf{K\textpm{}} or \textbf{M} fixed points. The former
set corresponds to the kination initial conditions and the latter
ones corresponds to a universe where matter dominates initially. In
both cases most trajectories are first attracted to the scaling fixed
point \textbf{Sc}. But because this fixed point is a saddle point
for $\kappa/\lambda>1$, eventually all the trajectories are repelled
and move to the de Sitter attractor \textbf{dS}. 

This represents the desirable sequence of events: initially the universe
is dominated by the kinetic energy of the scalar field, which is the
case for quintessential inflation models, or the matter component,
which is often encountered in the quintessence models. Next, the system
moves into the scaling fixed point, and for a long time the Universe
evolves with an effective equation of state that of the matter component. 

In the case of GR models and the exponential potential $V\left(\phi\right)$,
the scaling fixed point \textbf{Sc} is an attractor \citep{Copeland:1997et}.
That is, all the trajectories converge onto this point and remain
there. This is problematic, because the universe is not accelerating
at \textbf{Sc} in contrast to observations. The GB term, on the other
hand, converts this point into a saddle one and provides an escape
route. The scalar field eventually can come to dominate and cause
the universe to expand in an accelerated fashion in a new de Sitter
attractor point \textbf{dS}.

\todo[inline]{Remove this paragraph.}

\begin{figure}
\begin{centering}
\includegraphics[scale=0.4]{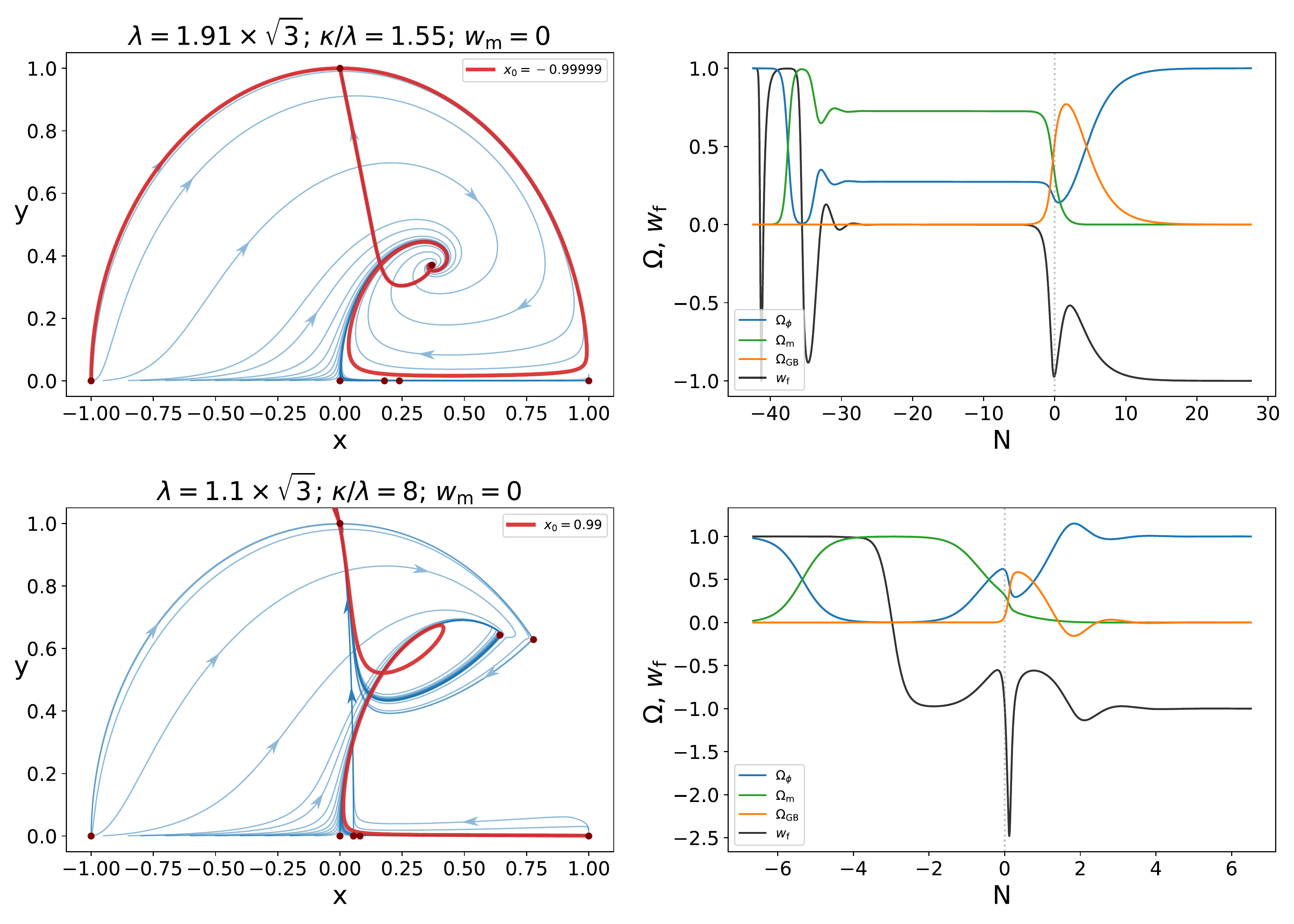}
\par\end{centering}
\caption{\label{fig:time-evolution}The time evolution of the density and the
effective equation of state for two selected trajectories (denoted
by red curves on the left column). They are selected such that $\Omega_{\mathrm{m}}$
and $w_{\mathrm{f}}$ are consistent with present day observations.}
\end{figure}

In Figure~\ref{fig:time-evolution} we show the time evolution of
the density parameters and the effective equation of state of the
``dark fluid''. The density parameters are defined \chngf{in eqs.~(\ref{=0003A9m})-(\ref{=0003A9GB}).
As discussed below eq.~(\ref{=0003A9const}), these parameters are
not bounded to lie at or bellow 1, which can be also seen in the bottom
right plot.} 

In order to apply the observational bounds on the equation of state
of Dark Energy we define a new parameter $\wf$. It can be interpreted
as the equation of state of an effective ``dark fluid'' that causes
the accelerated expansion. To do that eq.~(\ref{=0003F5H-GB}) can
be written as
\begin{eqnarray}
\eh & = & \frac{3}{2}\frac{\rho_{\phi}\left(1+\wf\right)+\rho_{\mathrm{m}}\left(1+w_{\mathrm{m}}\right)}{\rho_{\phi}+\rho_{\mathrm{m}}}\:,
\end{eqnarray}
where
\begin{eqnarray}
\wf & \equiv & \frac{P_{\phi}+8H^{2}\left(\ddot{\xi}-\dot{\xi}H\right)}{\rho_{\phi}}
\end{eqnarray}
and $\rho_{\phi}$ and $P_{\phi}$ are the energy and pressure densities
of the scalar field defined in eqs.~(\ref{=0003C1=0003D5}) and (\ref{P=0003D5}).
In terms of dimensionless variables in eq.~(\ref{dless-def}) the
last expression can be also written as
\begin{eqnarray}
\wf & = & -1+\frac{\frac{2}{3}\eh\left(1+2ux\right)-z^{2}\left(1+\wm\right)}{x^{2}+y^{2}}\:,
\end{eqnarray}
where $\eh$ is meant to be substituted with eq.~(\ref{eH-dyn}).

We next run a large number of numerical simulations of eqs.~(\ref{dx}),
(\ref{dy}) and (\ref{du-exp}) \chng{varying $\lambda$, $\kappa$
parameters as well as the initial conditions $x_{0}$ (but always
with $u_{0}=-10^{-25}$ and $\wm=0$)} and select those models which
have regions where $\Omega_{\mathrm{m}}=0.3147\pm0.0074$ and $\wf=-0.957\pm0.08$
\citep{Planck:2018vyg} are satisfied for the same value of $N$.
The scale factor is normalised such that $a=1$ ($N=0$) at that moment.

In Figure~\ref{fig:time-evolution} we show two of such models. On
the L.H.S. column we can see the phase portraits, where these models
(red curves) are drawn from and on the R.H.S. column we find the time
evolution of the density parameters and $\wf$. Both models have an
initial period of kination, which quickly gives way to the matter
domination . The model in the upper plot, has a long period of the
scaling behaviour before GB energy takes over. Eventually, all the
models settle down at the de Sitter attractor point, where $\Omega_{\phi}=1$
and all other $\Omega$'s vanish.

In the lower panel of Figure~\ref{fig:time-evolution} we can also
notice a quite generic feature of GB Dark Energy models, namely, that
for a brief period of time the effective equation of state of the
dark fluid can drop below $-1$.

\subsection{The Speed of Gravitational Waves}

Obviously, to select a realistic model of DE the consistency with
observational constraints on $\Omega_{\mathrm{m}}$ and $\wf$ parameters
is a necessary but not sufficient condition. There are some other
requirements that a viable model of cosmology must satisfy. Among
those requirements, especially in the case of GB model, is a negligibly
small deviation of the speed of gravitational waves $\cgw$ from the
speed of light.

As it was pointed out in \citep{Ezquiaga:2017ekz}, generically in
scalar-GB models $\cgw\ne1$. Due to the tight observational constraints
on $\cgw$ (see eq.~(\ref{=0003B1T-bound})) it was deemed that GBDE
models are excluded. However, such constraints do not fix $\cgw=1$,
they only place upper bounds on the deviation from $1$, albeit very
strong ones. And, as one could naively conclude from eq.~(\ref{bound-gen-num}),
that bound is not very constraining for GB models of DE. Moreover,
in GB models, $\cgw$ is not a constant, but varies with time. But
the constraint in eq.~(\ref{=0003B1T-bound}) applies only for a
short period over the history of the Universe.

To investigate this issue let us first write eq.~(\ref{=0003B1T})
in terms of dimensionless variables, defined in eq.~(\ref{dless-def})
\begin{eqnarray}
\alpha_{T} & = & \frac{\left(ux\right)'+\left(\eh-1\right)ux}{ux+1/2}=\frac{\Omega_{\mathrm{GB}}'+\left(\eh-1\right)\Omega_{\mathrm{GB}}}{\Omega_{\mathrm{GB}}-1}\:,\label{aT-diml}
\end{eqnarray}
where $\Omega_{\mathrm{GB}}$ is defined in eq.~(\ref{=0003A9GB}).
It might appear that $\alpha_{T}$ diverges at $ux=-1/2$, or equivalently
at \chngg{$\Omega_{\mathrm{GB}}=1$}. But, as can be seen from the
constraint equation (\ref{constr}), such a value is not allowed.

We can immediately notice from eq.~(\ref{aT-diml}) that $\alpha_{T}$
vanishes at the two fixed points, \textbf{Sc} and \textbf{dS}. These
fixed points are the most interesting ones. Unfortunately, if the
GB model is to be a good model of our Universe, the current stage
of the evolution cannot be represented by any of these two fixed points.
Instead, we should find ourselves somewhere on the trajectory between
\textbf{Sc} and \textbf{dS}, as it is also demonstrated in Figure~\ref{fig:time-evolution}.

\begin{figure}
\begin{centering}
\includegraphics[scale=0.55]{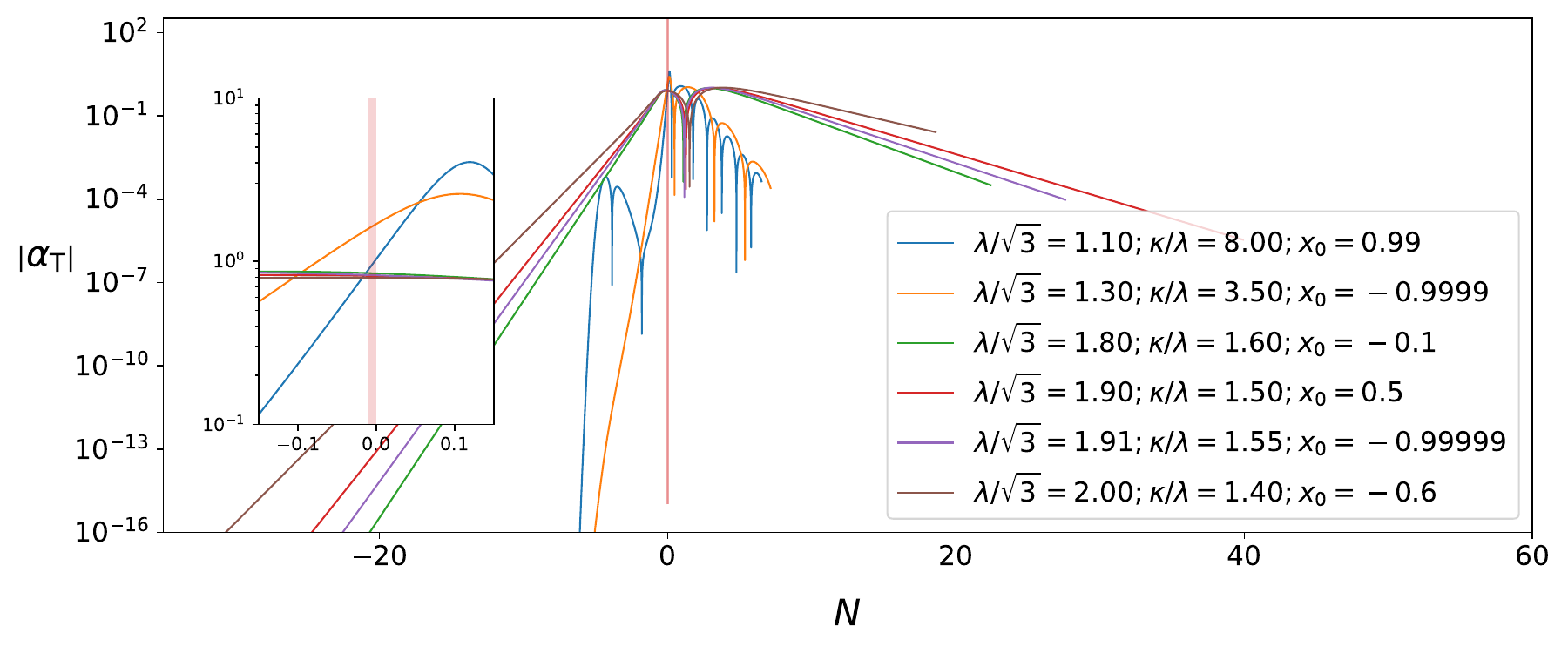}
\par\end{centering}
\caption{\label{fig:=0003B1T}The deviation of the speed of gravitational waves
from the speed of light (see eq.~(\ref{=0003B1T-def})) as a function
of time. $N=0$ represents the moment where the model predicts $\Omega_{\mathrm{m}}$
and $\protect\wf$ values consistent with present day observations.
This is also the moment where the bound in eq.~(\ref{=0003B1T-bound})
applies \chngf{(denoted by the very narrow vertical red rectangle)}.
Obviously all models violate this bound by many orders of magnitude.\chngf{The
inset zooms in the region where the curves cross the forbidden values.}}
\end{figure}

To see how $\alpha_{T}$ evolves with time we \chngf{ ran a large
number of simulations with $\lambda$ and $\kappa$ values in the
range $1/\sqrt{3}\le\lambda\le20\sqrt{3}$ and $1.1\le\kappa/\lambda\le20$.
In all of these simulations we took $w_{m}=0$ and the initial values
$y_{0}=10^{-3}$, $u_{0}=-10^{-25}$ and a range of $x_{0}$ values.
Among all the trajectories we selected the ones that satisfy the above
discussed bounds on $\Omega_{\mathrm{m}}$ and $\wf$.} Some of those
solutions are shown in Figure~\ref{fig:=0003B1T}. For clarity we
depict only a few of them. However, the ones displayed in Figure~\ref{fig:=0003B1T}
are representative of the whole set. \chngf{In particular, we always
observe that the maximum value of $\alpha_{T}$ is} very close to
$N=0$, exactly where the observational bounds in eq.~(\ref{=0003B1T-bound})
apply. \chngf{In the figure this bound is denoted by the red rectangle,
which resembles a vertical line due to its narrowness. The width of
this rectangle corresponds to $\Delta N\simeq0.0098$ \citep{Hjorth:2017yza}.
As can be seen in the inset of this figure, the maximum value of $\left|\alpha_{T}\right|$
today is always $\mathcal{O}\left(0.1\right)-\mathcal{O}\left(1\right)$,
which clearly falls outside the allowed range.} Hence, with high
confidence we can conclude that GBDE models with an exponential coupling
constant $\xi\left(\phi\right)$ are excluded by the observational
constraint on the speed of gravitational waves.

To get some insight why $\alpha_{T}$ is so large at this particular
moment, we can write eq.~(\ref{aT-diml}) in an alternative form
\begin{eqnarray}
\alpha_{T} & = & 2\left(\eh-3\right)+3\left(1-\wm\right)\frac{\Omega_{\mathrm{m}}}{1-\Omega_{\mathrm{GB}}}+6\frac{y^{2}}{1-\Omega_{\mathrm{GB}}}\:.\label{=0003B1T2}
\end{eqnarray}
This equality can be obtained either by using the dynamical equations
(\ref{dx})--(\ref{du}) to eliminate the time derivative from eq.~(\ref{aT-diml}),
or directly from eq.~(\ref{=0003F5H-GB}). In any case, it is important
to notice that this expression is very generic: it is valid for any
potential $V\left(\phi\right)$ and any GB coupling $\xi\left(\phi\right)$,
not necessarily exponential. At $N=0$ observations require $\eh\sim0.5$.
Hence, the absolute value of the first term is of order $\sim1$.
Observations also require $\wm=0$ and $\Omega_{\mathrm{m}}\sim0.3$.
We can use the latter in the constraint in eq.~(\ref{=0003A9const})
to write $\Omega_{\phi}+\Omega_{\mathrm{GB}}\sim0.7$. Moreover, between
the scaling fixed point \textbf{Sc} and the de Sitter one \textbf{dS},
the GB density parameter $\Omega_{\mathrm{GB}}>0$. Therefore, $0<\Omega_{\mathrm{GB}}<0.7$
and the second term in eq.~(\ref{=0003B1T2}) must be of order $\mathcal{O}\left(0.1\right)$
to $\mathcal{O}\left(1\right)$. Finally $y^{2}\le\Omega_{\phi}<0.7$
and the last term of order $\mathcal{O}\left(1\right)$ at the maximum.
Barring precise cancellations, this leads to the conclusion that generically
$\left|\alpha_{T}\right|\sim\mathcal{O}\left(0.1\right)-\mathcal{O}\left(1\right)$.\footnote{The observation that $\Omega_{\mathrm{m}}\sim0.1$ quite generically
leads to $\left|\alpha_{T}\right|\gtrsim0.1$ raises another question.
As it follows from eq.~(\ref{bound-gen-num}), such a large value
of $\left|\alpha_{T}\right|$ implies a very large change in the GB
coupling. But one has to remember that the GB term is only the lowest
order term in the series of low-energy effective string theory corrections
\citep{Boulware:1985wk,Bento:1995qc,Sami:2005zc}. Since $\xi$ varies
so much, one must wonder if it is consistent to neglect the higher
order corrections over the whole range of the evolution.}

\section{The Linear Gauss-Bonnet Coupling\label{sec:lin}}

As we have seen above, models with an exponential GB coupling could
potentially provide a reasonable history of the Universe and explain
DE. Unfortunately, such models predict the speed of GW in the current
Universe that violates the allowed values by many orders of magnitude.
In this section we investigate another option: the linear GB coupling
\begin{eqnarray}
\xi & \propto & \phi\:.\label{=0003BElin}
\end{eqnarray}
As it is shown in the Appendix, such a dynamical system has the same
fixed points as the $\ehc\uc=0$ subset in Table~\ref{tab:pc-gen}. 

To study the stability of those fixed points, we can linearise eqs.~(\ref{dx})--(\ref{du})
and use $\xi_{,\phi\phi}=0$. Equivalently, we can use the results
in section~\ref{sec:Exponential} by setting $\kappa=0$. This gives
the eigenvalues at the \textbf{Sc} point 
\begin{eqnarray}
m_{1} & = & -3\left(1+\wm\right)\:,\\
m_{\pm} & = & -\frac{1}{2}\left(3-\ehc\right)\left[1\pm\sqrt{1-\frac{8\ehc}{3-\ehc}\cdot\left(1-2\frac{\ehc}{\lambda^{2}}\right)}\right]\:.
\end{eqnarray}
As discussed in section~\ref{sec:Exponential} the real part of $m_{\pm}$
eigenvalues are always negative. We can also see from the first equation
above that $m_{1}$ is negative too. Hence, this fixed point is always
an attractor.

On the other hand, the eigenvalues at the de Sitter fixed point \textbf{dS}
are
\begin{eqnarray}
m_{1} & = & -\frac{3}{2}\left(1+\wm\right)\:,\\
m_{\pm} & = & -\frac{3}{2}\left[1\pm\sqrt{1+\frac{8\lambda^{2}}{3\left(2+3\lambda^{2}\right)}}\right]\:.
\end{eqnarray}
It is clear that $m_{1},\:m_{+}<0$ and $m_{-}>0$. Hence, this fixed
point is a saddle.

\begin{figure}
\begin{centering}
\includegraphics[scale=0.3]{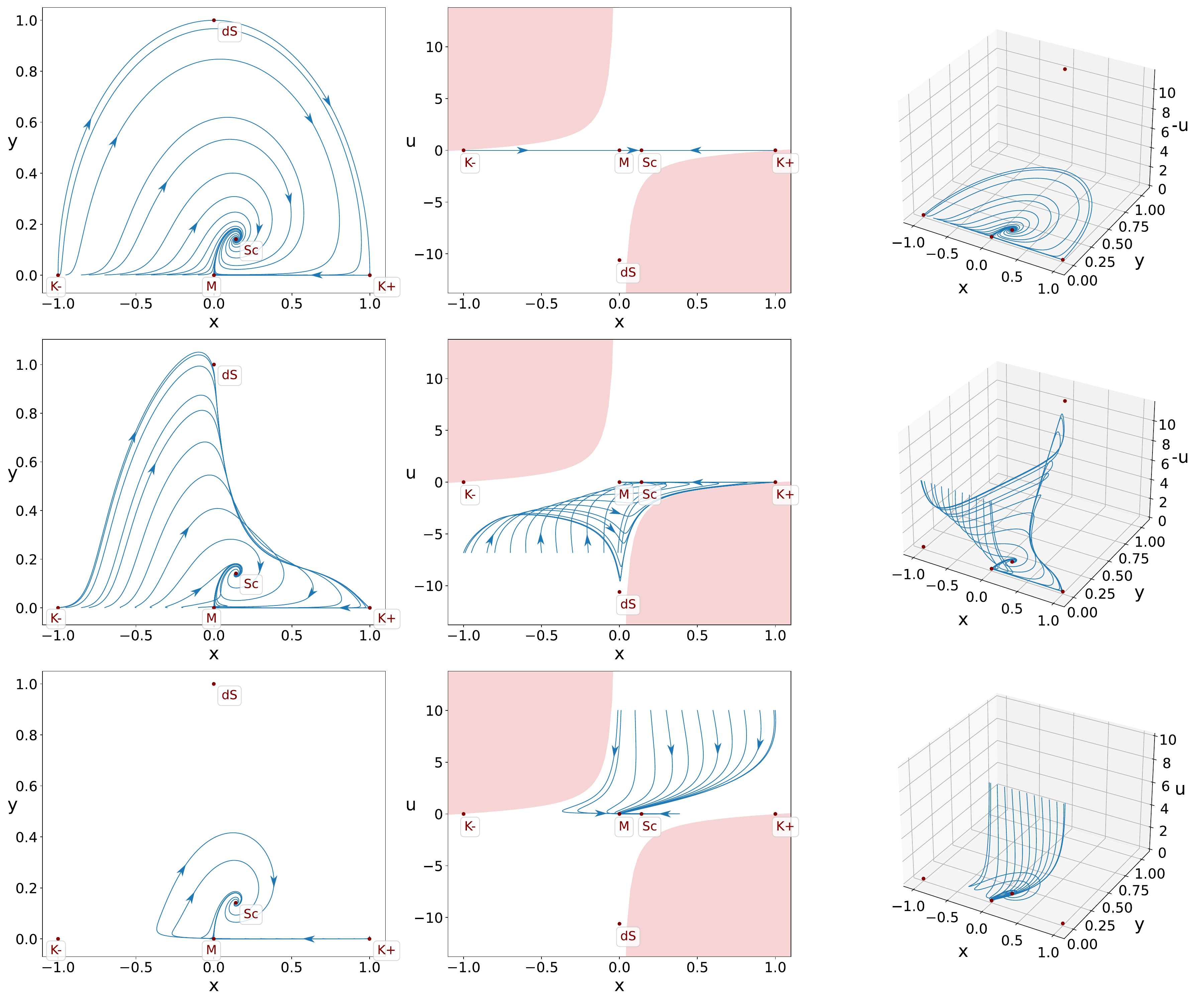}
\par\end{centering}
\caption{\label{fig:phlin}The phase space of the linear model, with $\xi\left(\phi\right)$
defined in eq.~(\ref{=0003BElin}). The notation and the red shaded
regions are the same as in Figure~\ref{fig:portraits}. All three
rows correspond to the same model with $\lambda=5\sqrt{3}$ and $\protect\wm=0$,
but different initial conditions of $u$. In the first row $u_{0}=-10^{-25}$.
As we can see the evolution remains within the $u=0$ plane. To demonstrate
that the scaling fixed point \textbf{Sc} is an absolute attractor
we also run simulations with large initial values $\left|u_{0}\right|$.
In the second row we can see that some trajectories are initially
attracted towards the de Sitter fixed point \textbf{dS}, but since
for a linear $\xi\left(\phi\right)$ this point is a saddle, eventually
all trajectories converge onto \textbf{Sc}. Notice that in the last
column we plot $-u$ on the vertical axis of the first two plots and
\chngf{$+u$} of the third plot.}

\end{figure}

What the above result shows is that a realistic cosmological scenario
is impossible with the linear GB coupling. The scaling fixed point
is an attractor and there are no solutions which display a long matter
dominated period followed by an accelerated expansion. This is also
demonstrated in Figure~\ref{fig:phlin}.

One might wish to generalise the analysis presented here to higher
order polynomial \chngf{ or other functions $\xi\left(\phi\right)$
as, for example, considered in refs.~\citep{Silva:2017uqg,Doneva:2017duq}
in the context of astrophysical compact objects. Unfortunately, other
functional forms of $\xi\left(\phi\right)$ are not amenable to the
presented methods of analysis, as the equations become non-self-similar.
Therefore, the relevance of other $\xi\left(\phi\right)$ functions
in explaining Dark Energy is left for future investigations.}

\section{The Case of $\alpha_{T}=0$\label{sec:aT=00003D0}}

As can be seen from the expression of $\alpha_{T}$ in eq.~(\ref{=0003B1T})
this parameter can be made to vanish if one arranges for the GB coupling
in such a way that its time evolution obeys the condition in eq.~(\ref{bound-spec}).
Unfortunately, this condition does not provide us with a functional
form of $\xi\left(\phi\right)$. Nevertheless, it provides with enough
information to investigate the relevant aspects of such a dynamical
system.

First, notice that eq.~(\ref{bound-spec}), written in terms the
dimensionless variables in eq.~(\ref{dless-def}), becomes
\begin{eqnarray}
24H^{2}\xi_{,\phi\phi}x^{2} & = & u\left[\left(\eh+1\right)x-x'\right]\,.\label{aT0}
\end{eqnarray}
It provides an additional constraint that can be used to eliminate
the $\xi_{,\phi\phi}$ term in eq.~(\ref{du1}), at least for fixed
points with $\xc\ne0$. Otherwise the LHS of the above equation vanishes
in any case. The fixed points are computed in the Appendix and summarised
in Table~\ref{tab:=0003B1T=00003D0}. As one would expect, all the
fixed points with $\uc\ehc=0$ are the same as in Table~\ref{tab:pc-gen}.

To determine the stability of fixed points we again linearise eqs.~(\ref{dx})--(\ref{du})
and also eq.~(\ref{aT0}). At the de Sitter fixed point \textbf{dS}
this linear system reduces to
\begin{eqnarray}
\delta x' & = & \delta x\,,\\
\delta u' & = & \left(4+3\lambda^{2}\right)\delta x\,,\\
\delta z' & = & -\frac{3}{2}\left(1+\wm\right)\delta z\,,
\end{eqnarray}
where we used that the linearised eq.~(\ref{aT0}) at \textbf{dS}
implies the constraint $\delta x'=\delta x$. Combined with the linearised
eqs.~(\ref{dx}) and (\ref{constr}) we find $\delta u=\left(4+3\lambda^{2}\right)\delta x$.
It is easy to compute the eigenvalues, which are
\begin{eqnarray}
m_{1} & = & 0\:,\\
m_{2} & = & 1\:,\\
m_{3} & = & -\frac{3}{2}\left(1+\wm\right)\:.
\end{eqnarray}
We can see that the last two eigenvalues have opposite signs. Hence,
in this case \textbf{dS} is a saddle fixed point, no longer an attractor.

In the neighbourhood of the scaling fixed point \textbf{Sc}, the linear
equation for $\delta u$ takes the form
\begin{eqnarray}
\delta u' & = & \left(1-\ehc\right)\delta u\:,
\end{eqnarray}
where $1-\ehc=-\frac{1}{2}\left(1+3\wm\right)$ is always negative
within the range $0\le\wm<1$. Therefore, in the $u$ direction of
the phase space this fixed point is attractive and there are no trajectories
that flow from \textbf{Sc} to \textbf{dS}.

In summary, the condition $c_{\mathrm{GW}}=1$ implies the scaling
fixed point \textbf{Sc} to be an attractor, just as in models of GR
\citep{Copeland:1997et} and \textbf{dS} becomes a saddle point. This
leads to the conclusion that there are no solutions which reproduce
a long, matter-like domination period and asymptotically approach
the de Sitter solution, which is required to reproduce the evolution
of the Universe.

\begin{table}
\begin{centering}
\begin{tabular}{|c|c|c|c|c|c|}
\cline{2-6} \cline{3-6} \cline{4-6} \cline{5-6} \cline{6-6} 
\multicolumn{1}{c|}{} & $\xc$ & $\yc$ & $\uc$ & $\zc$ & Expansion rate $\ehc$\tabularnewline
\hline 
\textbf{M} & 0 & 0 & 0 & 1 & $\frac{3}{2}\left(1+\wm\right)$\tabularnewline
\hline 
\textbf{K\textpm{}} & $\pm1$ & 0 & 0 & 0 & $3$\tabularnewline
\hline 
\textbf{I} & $\frac{\lambda}{\sqrt{6}}$ & $\sqrt{1-\frac{\lambda^{2}}{6}}$ & 0 & 0 & $\frac{1}{2}\lambda^{2}$\tabularnewline
\hline 
\textbf{Sc} & $\sqrt{\frac{3}{2}}\frac{1+\wm}{\lambda}$ & $\frac{\sqrt{\frac{3}{2}\left(1-\wm^{2}\right)}}{\lambda}$ & 0 & $\sqrt{1-\frac{3\left(1+\wm\right)}{\lambda^{2}}}$ & $\frac{3}{2}\left(1+\wm\right)$\tabularnewline
\hline 
\textbf{dS} & 0 & $1$ & $-\sqrt{\frac{3}{2}}\lambda$ & 0 & 0\tabularnewline
\hline 
\textbf{$\overline{\text{\textbf{IV}}}$} & $\sqrt{\frac{2}{3}}\frac{1}{\lambda}$ & $\frac{2}{\sqrt{3}\lambda}$ & $\sqrt{\frac{3}{2}}\frac{1}{\lambda}\left(1-\frac{\lambda^{2}}{2}\right)$ & 0 & 1\tabularnewline
\hline 
\end{tabular}
\par\end{centering}
\caption{\label{tab:=0003B1T=00003D0}Fixed points of the dynamical system
that satisfies $\alpha_{T}=0$.}
\end{table}

\section{Summary and Conclusions}

In this work we investigate Gauss-Bonnet Dark Energy (GBDE) models.
Generically such models predict the speed of gravitational waves different
from the speed of light. In view of the tight observational constraints
on such deviations, denoted by $\at$ (see eq.~(\ref{=0003B1T-bound})),
such models are considered to be disfavoured. However, the deviation
is time dependent and the bound, although tight, is an upper bound,
which is only applicable for very late Universe. Hence, before excluding
GBDE models we need to perform a more detailed analysis. Moreover,
if the bound in eq.~(\ref{=0003B1T-bound}) is expressed in terms
of the variation of the GB coupling function $\xi\left(\phi\right)$,
it might appear to be a \deld weak bound, as shown in eq.~(\ref{bound-gen-num}).

To see if GBDE models can indeed survive the observational constraints
on $\at$ we perform the dynamical \chngd{systems} analysis. We
assume that the scalar field has an exponential potential and find
that such a dynamical system quite generically has the scaling and
de Sitter fixed points (among others), denoted by \textbf{Sc} and
\textbf{dS} respectively in this work. 

At \textbf{Sc} the scalar field adjusts in such a way that it mimics
the behaviour of the \chngd{background} matter component. In particular,
the equation of state of \chngd{the scalar field} is the same as
that of the matter component. In the case of the exponential GB coupling,
depending on the magnitude of the exponent, \textbf{Sc} is not stable:
it is a saddle point. This is in contrast to the General Relativistic
quintessence models with an exponential potential \citep{Copeland:1997et}.
In the latter setup the scaling fixed point is an attractor, which
makes it impossible to use as an explanation for the accelerated expansion
of the Universe. In GBDE \textbf{Sc }can have an unstable direction
which links to \textbf{dS}, the latter being an attractor. Using numerical
solutions we show that if the universe starts with a very small GB
term and it is either kination or matter dominated, initially all
solutions evolve towards the \textbf{Sc} fixed point. They linger
in the neighbourhood of \textbf{Sc} for a long period of time and
eventually change its course towards \textbf{dS}. This behaviour is
beneficial to modelling quintessential inflation, because it allows,
without extreme fine-tuning, to bridge the enormous energy density
gap (110 orders of magnitude) between \chngg{inflation} and dark
energy. We find numerous models that follow this scenario and can
predict observationally allowed values for the matter energy density
and DE equation of state.

Moreover, at \textbf{Sc }and \textbf{dS }fixed points the speed of
gravitational waves is exactly the same as that of the speed of light,
i.e. $\at=0$. Unfortunately, if this model is to represent the evolution
of the actual Universe, we cannot be living either on the scaling
or de Sitter fixed points, but somewhere in between. However, as we
show in Figure~\ref{fig:=0003B1T}, $\left|\at\right|$ changes from
$0$ to $\sim1$ in between these fixed points. Moreover, as we argue
below eq.~(\ref{=0003B1T2}), if $\Omega_{\mathrm{m}}\sim0.1$, which
corresponds to the current value, quite generically one expects $\left|\at\right|\sim\mathcal{O}\left(0.1\right)-\mathcal{O}\left(1\right)$,
which is ruled out by many orders of magnitude. We conclude that the
bound in eq.~(\ref{=0003B1T-bound}) makes GBDE with an exponential
$\xi\left(\phi\right)$ function unviable.

In view of the above conclusion\chngd{,} we investigated \deld
other choices of $\xi\left(\phi\right)$. First, we demonstrate that
for a linear function $\xi\left(\phi\right)$ the de Sitter fixed
point is a saddle point, and the Scaling fixed point becomes an attractor.
This makes it impossible to find any viable solution that would be
consistent with the evolution of the Universe. We cannot apply our
method to study more generic functions $\xi\left(\phi\right)$, \chngf{such
as monomials or steeper than exponential potentials studied, for example,
in Refs.~\citep{Silva:2017uqg,Doneva:2017duq}. This is because dynamical
equations lose their self-similar character.}

The advantage of the above described analysis is that we have an explicit
$\xi\left(\phi\right)$ function. But as we showed, this does not
provide a viable cosmological solution. Another hope to make GBDE
models conform to observational constraints is to impose the condition
$\at=0$, as can be seen in eq.~(\ref{=0003B1T}). In this case we
lose the benefit of having an explicit functional form of $\xi\left(\phi\right)$,
but we gain an additional constraint equation (\ref{aT0}). As it
is demonstrated in Section~\ref{sec:aT=00003D0}, the stability of
fixed points in this case is very similar to the linear model. That
is, there are no solutions which provide a long period of matter domination
followed by an accelerated expansion.

In summary, we find that a GBDE model with an exponential scalar field
potential and an exponential GB coupling function could provide a
realistic model of DE. However, the recent bounds on the speed of
GW rules out this possibility by many orders of magnitude. If, on
the other hand, we look for models that do satisfy $\at=0$, then
it is impossible to find a scenario consistent with other cosmological
observations.

The negative conclusions reached in this work apply to the metric
formulation of the Gauss-Bonnet model. But we know that some modified
gravity models that violates the $\at$ bound in eq.~(\ref{=0003B1T-bound})
become viable again in the Palatini formalism \citep{Kubota:2020ehu}.
One can hope that a similar modification could save the GBDE model
too.\footnote{Analogously, if the GB term is coming from the breaking of the Weyl
symmetry, also a Weyl term should be added to the action (see for
instance Refs.~\citep{Coriano:2022ftl,Fernandes:2022zrq}).} We intend to study this possibility in future publications.

\begin{acknowledgments}
M.K. is supported by the Mar\'{\i}a Zambrano grant, provided by the
Ministry of Universities from the Next Generation funds of the European
Union. This work is also partially supported by the MICINN (Spain)
projects PID2019-107394GB-I00/AEI/10.13039/501100011033 (AEI/FEDER,
UE). K.D. was supported, in part, by the Lancaster-Manchester-Sheffield
Consortium for Fundamental Physics under STFC grant: ST/T001038/1.A.R.
is supported by the Estonian Research Council grant PRG1055. 

For the purpose of open access, the authors have applied a Creative
Commons Attribution (CC BY) license to any Author Accepted Manuscript
version arising.
\end{acknowledgments}

\appendix

\section{The Fixed Points \label{sec:The-Fixed-Points}}

In this section we derive fixed points of the dynamical system in
eqs.~(\ref{dx})--(\ref{constr}) that are summarised in Table~\ref{tab:pc-gen}.
Such points, or higher dimension structures\chngf{,} are regions
of the phase space where $x'=y'=z'=0$. We denote the constant values
of $x$, $y$, $z$ at these regions by $\xc$, $\yc$, $\zc$. \emph{A
priori} we do not impose the condition $u'=0$ at fixed points, but
it follows from the equations. In order to see that we can take the
derivative of the constraint equation with respect to $N$. At a fixed
point this gives
\begin{eqnarray}
\left.u'\right|_{x=\xc}\xc & = & 0\,.\label{uc-cond}
\end{eqnarray}
This equation allows for $u'\ne0$ if $\xc=0$. However, if we plug
$x'=\xc=0$ into eq.~(\ref{dx}) and (\ref{eH-dyn}) we find
\begin{eqnarray}
0 & = & \sqrt{\frac{3}{2}}\lambda\yc^{2}+u\left(1-\ehc\right)\:,
\end{eqnarray}
where, in this case, $\ehc$ is the value of $\eh$ in eq.~(\ref{eH-dyn})
at $x'=y'=z'=\xc=0$. As it is clear from the above equation, all
quantities on the RHS are constant but $u$. Hence, $u=\uc$ must
be also a constant even for $\xc=0$. In summary, the algebraic equations
to find fixed point values are given by
\begin{eqnarray}
0 & = & \left(\ehc-3\right)\xc+\sqrt{\frac{3}{2}}\lambda\yc^{2}+\uc\left(1-\ehc\right)\:,\\
0 & = & \left(\ehc-\sqrt{\frac{3}{2}}\lambda\xc\right)\yc\:,\\
0 & = & \left[\ehc-\frac{3}{2}\left(1+\wm\right)\right]\zc\:,
\end{eqnarray}
where
\begin{eqnarray}
\ehc & = & \left[3\xc^{2}+\frac{3}{2}\left(1+\wm\right)\zc^{2}-\uc\xc\right]\frac{1}{1+\uc\xc}
\end{eqnarray}
and all the values have to satisfy the constraint in eq.~(\ref{constr}).
There are four possible solutions of this system:
\begin{description}
\item [{A}] $\left(x,y,u\right)=\left(\xc,0,\frac{3\left(\wm-1\right)}{3\wm+1}\xc\right)$,
which is valid for any value of $\xc$. The expansion rate in this
case is given by $\ehc=\frac{3}{2}\left(1+\wm\right)$. We can see
that this is the scaling solution, where the scalar field adjusts
to the equation of state of the matter component.
\item [{B}] $\left(x,y,u\right)=\left(\xc,0,\frac{\xc^{2}-1}{2\xc}\right)$
with $\ehc=\frac{5\xc^{2}+1}{\xc^{2}+1}$
\item [{C}] $\left(x,y,u\right)=\left(\xc,\sqrt{1-\xc^{2}+2\xc\frac{3\xc-\sqrt{\frac{3}{2}}\lambda}{1+\sqrt{\frac{3}{2}}\lambda\xc}},\frac{3\xc-\sqrt{\frac{3}{2}}\lambda}{1+\sqrt{\frac{3}{2}}\lambda\xc}\right)$
with $\ehc=\sqrt{\frac{3}{2}}\lambda\xc$
\item [{D}] $\left(x,y,u\right)=\left(\sqrt{\frac{3}{2}}\frac{1+\wm}{\lambda},\frac{\sqrt{\frac{3}{2}\left(1-\wm^{2}\right)+\frac{\lambda}{\sqrt{6}}\left(1+3\wm\right)\uc}}{\lambda},\uc\right)$
with $\ehc=\frac{3}{2}\left(1+\wm\right)$. This is a second scaling
solution, which is valid for any (allowed by the constraint) value
of $\uc$.
\end{description}
As we can see, these ``fixed points'' are actually curves in the
three dimensional phase space. In the case of \textbf{A} to \textbf{C},
the curves are parametrised by the $\xc$ value, which can vary within
the range where $y$, $u$ and $z$ remain real. In the case of \textbf{D},
the value of $x$ is fixed to $\xc=\sqrt{\frac{3}{2}}\frac{1+\wm}{\lambda}$,
but $u$ is free to vary within the similarly defined range.

We can infer more about the structure of fixed points if we use the
definition $u$ in eq.~(\ref{dless-def}). Taking the derivative
of that expression we get eq.~(\ref{du}), which we rewrite it here
for convenience:
\begin{eqnarray}
u' & = & -2\eh u+24H^{2}\xi_{,\phi\phi}x\:.\label{du1}
\end{eqnarray}
Hence, at fixed points we get the relation
\begin{eqnarray}
12H^{2}\xi_{,\phi\phi}\xc & = & \ehc\uc\:.\label{du1-fp}
\end{eqnarray}

If we take an exponential GB function, given in eq.~(\ref{=0003BE=0003BA}),
the above equation becomes
\begin{eqnarray}
\sqrt{\frac{3}{2}}\kappa\xc\uc & = & \ehc\uc\:.
\end{eqnarray}
Note, that we did not cancel $\uc$ factors, because $\uc=0$ is an
allowed solution. Plugging various values, that are consistent with
this equation, into the system \textbf{A}--\textbf{D}, we obtain
fixed points that are summarised in Table~\ref{tab:pc-gen}. In that
table $\xc=\beta$ at the fixed point \textbf{G}. This is a solution
of the cubic equation, which is given by
\begin{eqnarray}
\beta & \equiv & \frac{1}{3\sqrt{3}\kappa}\left(5\sqrt{2}+\frac{9\kappa^{2}-50}{\alpha^{1/3}}-\alpha^{1/3}\right)\:,\label{beta-def}
\end{eqnarray}
where
\begin{eqnarray}
\alpha & \equiv & \sqrt{2}\left(27\kappa^{2}-250\right)+\sqrt{2\left(27\kappa^{2}-250\right)^{2}+\left(9\kappa^{2}-50\right)^{3}}
\end{eqnarray}
and $\kappa$ is an exponent of $\xi\left(\phi\right)$.

It is interesting to note that all fixed points with $\uc\ehc\ne0$
(\textbf{S2}, \textbf{G}, \textbf{S3} and \textbf{IV}) exist only
if the GB function $\xi\left(\phi\right)$ is of the form
\begin{eqnarray}
\xi\left(\phi\right) & = & c_{1}\frac{\phi}{\mpl}+c_{2}\mathrm{e}^{\kappa\frac{\phi}{\mpl}}\:,
\end{eqnarray}
where $c_{2}\ne0$ and $\kappa=\sqrt{2/3}\cdot\ehc/\xc$. Or, more
precisely, it is sufficient that the GB function asymptotically approaches
this solution, as the trajectory in the phase space gets closer to
those fixed points. This result can be obtained by integrating eq.~(\ref{du1-fp})
and using the fact that for $\xc\ehc\ne0$ we have
\begin{eqnarray}
H & = & H_{0}\mathrm{e}^{-\frac{\ehc}{\sqrt{6}\xc}\frac{\phi}{\mpl}}\:.
\end{eqnarray}

For the linear $\xi\left(\phi\right)$, i.e. for $\xi_{,\phi\phi}=0$,
it follows from eq.~(\ref{du1-fp}) that only fixed points with $\ehc\uc=0$
are present. Their values are summarised in Table~\ref{tab:pc-gen}. 

In the case of $\alpha_{T}=0$ constraint, eq.~(\ref{aT0}) at a
fixed point can be written as 
\begin{eqnarray}
24\left.H^{2}\xi_{,\phi\phi}\right|_{\mathrm{c}}\xc^{2} & = & \left(\ehc+1\right)\xc\uc\:.\label{aT-fp}
\end{eqnarray}
For fixed points with $\xc=0$ this equation vanishes identically.
Looking at the system \textbf{A}--\textbf{D}, we see that only two
such fixed points exist: $\left(x,y,u\right)=\left(0,0,0\right)$
and $\left(x,y,u\right)=\left(0,1,-\sqrt{\frac{3}{2}}\lambda\right)$.
These correspond to points \textbf{M} and \textbf{dS} in Table~\ref{tab:pc-gen}.
For $\xc\ne0$ we can equate eq.~(\ref{du1-fp}) with (\ref{aT-fp})
and get
\begin{eqnarray}
2\ehc\uc & = & \left(\ehc+1\right)\uc\:.
\end{eqnarray}
For $\uc=0$ the equation vanishes identically and we obtain the same
fixed points as $\uc=0$ points in Table~\ref{tab:pc-gen}. On the
other hand, there is only one fixed point that satisfies $\uc\ne0$.
That fixed point must have $\ehc=1$. Looking at the system \textbf{A}--\textbf{D},
we can see that only one such point is allowed, which is the relation
\textbf{C} with $\left(x,y,u\right)=\left(\sqrt{\frac{2}{3}}\frac{1}{\lambda},\frac{2}{\sqrt{3}\lambda},\sqrt{\frac{3}{2}}\frac{1}{\lambda}\left(1-\frac{\lambda^{2}}{2}\right)\right)$.
All these fixed points are summarised in Table~\ref{tab:=0003B1T=00003D0}.

\begin{figure}
\begin{centering}
\includegraphics[scale=0.55]{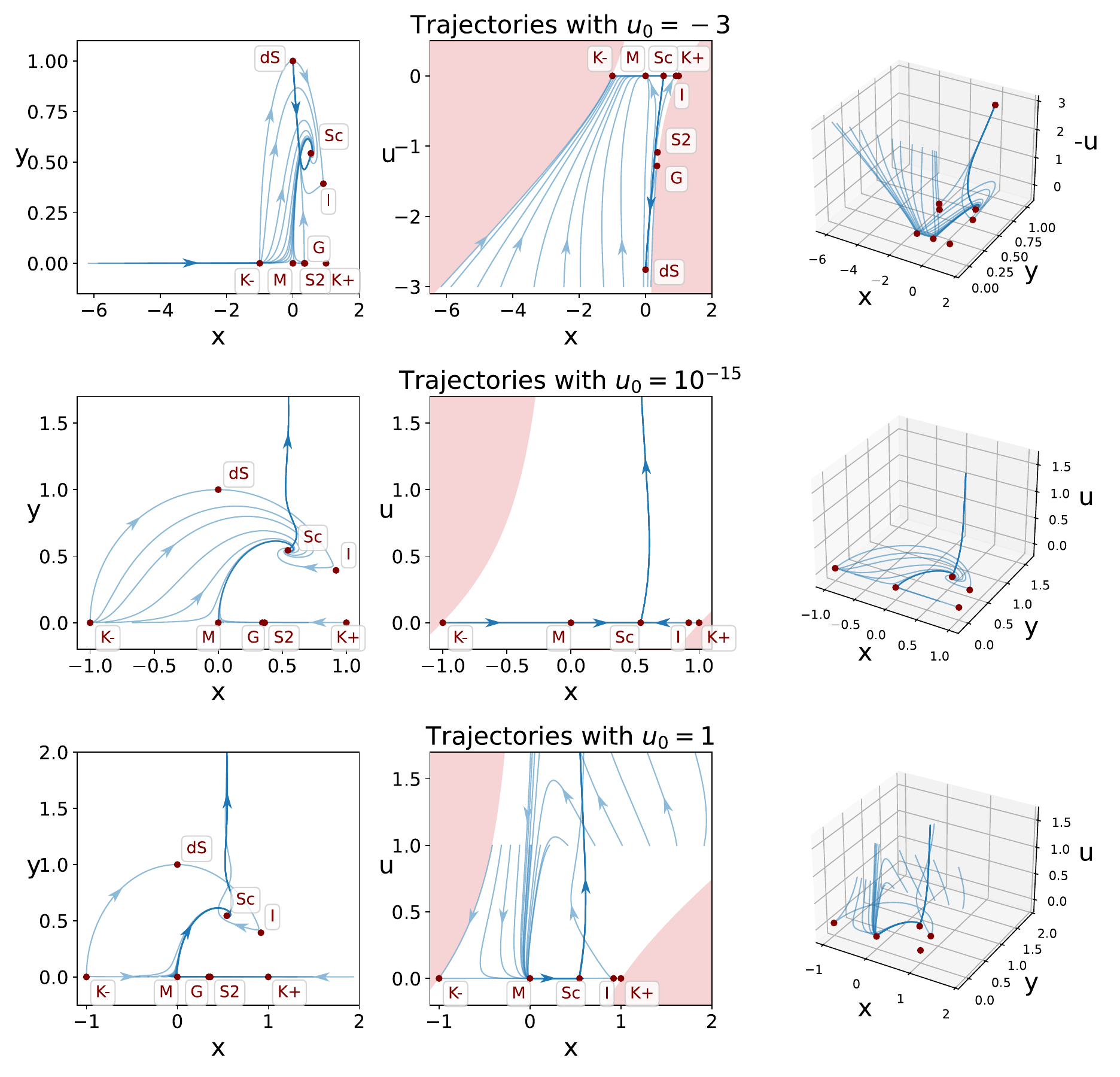}
\par\end{centering}
\caption{\label{fig:phpt-u}A few examples of phase portraits with different
initial values $u_{0}$. The notation, including the red shaded regions,
is the same as in Figure~\ref{fig:portraits}. \chngg{All models
correspond to the same value $\lambda=1.3\times\sqrt{3}$, $\kappa/\lambda=3/2$
and $w_{m}=0$. These values are also used in the upper row of Fig.~\ref{fig:portraits}.
Notice that the vertical axis of the upper right column is $-u$.}}
\end{figure}

\section{\label{sec:Phase-Portraits-u}Phase Portraits with More Generic Initial
Values}

\chngf{In Fig.~\ref{fig:portraits} we show the phase portrait for
the exponential GB coupling with two sets of values of $\lambda$
and $\kappa$. All displayed trajectories start with the initial value
$u_{0}=-10^{-25}$ in that figure. The absolute value of $u_{0}$
is chosen to make the GB term completely negligible initially. This
choice is partly motivated by simplicity, because, as can be seen
in Fig.~\ref{fig:phpt-u}, the dynamics between the scaling fixed
point \textbf{Sc} and the de Sitter one \textbf{dS} is not affected
much, at least for some part of the trajectories. Small initial $u_{0}$
values are also motivated by the model in ref.~\citep{vandeBruck:2017voa},
where the GB term does not affect the early dynamics of the universe.
The sign of $u_{0}$, on the other hand, is crucially important: only
those trajectories that satisfy $u_{0}<0$ can eventually be attracted
towards \textbf{dS} fixed point.\textbf{ }We find that the sign of
$u$ is preserved by the evolution equations.

To demonstrate all these points, we provide several plots in Fig.~\ref{fig:phpt-u}
with numerically integrated trajectories for several $u_{0}$ values.
For clarity we display plots with negative and positive initial $u$
values separately. Also note that we choose the value of $\kappa$
that is close to $\lambda$. If the ratio $\kappa/\lambda$ is too
large, the attraction towards the \textbf{Sc} fixed point becomes
weak, and most of the trajectories do not pass close to this point.
This observation is valid for positive as well as negative $u_{0}$
initial conditions.}\delg

\bibliographystyle{aipnum4-2}
\bibliography{gbquint.bbl}

\end{document}